\documentclass[10pt,conference]{IEEEtran}
\makeatletter
\def\ps@headings{%
\def\@oddhead{\mbox{}\scriptsize\rightmark \hfil \thepage}%
\def\@evenhead{\scriptsize\thepage \hfil \leftmark\mbox{}}%
\def\@oddfoot{}%
\def\@evenfoot{}}
\makeatother
\usepackage{cite}
\usepackage{amsmath,amssymb,amsfonts}
\usepackage{algorithmic}
\usepackage{graphicx}
\usepackage{textcomp}
\usepackage{xcolor}

\usepackage{catchfile} 
\usepackage{multicol} 
\usepackage{multirow} 
\usepackage{listings} 
\usepackage{scalerel} 
\usepackage{subcaption} 

\def\BibTeX{{\rm B\kern-.05em{\sc i\kern-.025em b}\kern-.08em
    T\kern-.1667em\lower.7ex\hbox{E}\kern-.125emX}}

\newcommand{\theos}{Contiki NG}
\newcommand{\thename}{BLEND}

\IEEEoverridecommandlockouts
\IEEEpubid{\begin{tabular}[t]{@{}l@{}}
\copyright{}2022 IEEE. Personal use of this material is permitted. Permission from \\
IEEE must be obtained for all other uses, in any current or future media,\\
including reprinting/republishing this material for advertising or promotio- \\
nal purposes, creating new collective works, for resale or redistribution to\\
servers or lists, or reuse of any copyrighted component of this work in\\ 
other works.\hfill\end{tabular}
\hspace{\columnsep}\makebox[\columnwidth]{ }} 

\begin{document}

\title{\thename: Efficient and blended IoT data storage and communication with application layer security} 

\author{\IEEEauthorblockN{Joel H\"oglund, Shahid Raza}
\IEEEauthorblockA{RISE Research Institutes of Sweden\\
Isafjordsgatan 22, 16440 Kista, Stockholm \\
\{joel.hoglund, shahid.raza\}@ri.se}
}

\maketitle

\begin{abstract}
Many IoT use cases demand both secure storage and secure communication. Resource-constrained devices cannot afford having one set of crypto protocols for storage and another for communication. Lightweight application layer security standards are being developed for IoT communication. Extending these protocols for secure storage can significantly reduce communication latency and local processing.
 
We present BLEND, combining secure storage and communication by storing IoT data as pre-computed encrypted network packets. Unlike local methods, BLEND not only eliminates separate crypto for secure storage needs, but also eliminates a need for real-time crypto operations, reducing the communication latency significantly. Our evaluation shows that compared with a local solution, BLEND reduces send latency from 630 $\mu$$s$ to 110 $\mu$$s$ per packet. BLEND enables PKI based key management while being sufficiently lightweight for IoT. BLEND doesn't need modifications to communication standards used when extended for secure storage, and can therefore preserve underlying protocols’ security guarantees.
\end{abstract}

\begin{IEEEkeywords}
Secure storage, communication security, application layer security, OSCORE, EDHOC, IoT
\end{IEEEkeywords}

\section{Introduction}
\label{sec:intro}
IoT is being deployed in extremely heterogeneous and wild scenarios such as agriculture monitoring, battlefields, remote surveillance, power-line monitoring, flood monitoring, and telemedicine. Most of these deployments require data confidentiality and/or integrity while at rest as well as in transit. While traditional Datagram TLS (DTLS)~\cite{RFC6347-dtls} has been extended to IoT, it is still too heavy for many IoT scenarios and lack full end-to-end security across different transport layer technologies. New Application layer protocols, namely OSCORE~\cite{RFC8613-oscore} and EDHOC~\cite{EDHOC}, are specifically designed for resource-constrained IoT and offer full end-to-end security. 

\begin{figure}[t]
\centering
\includegraphics[width=\columnwidth]{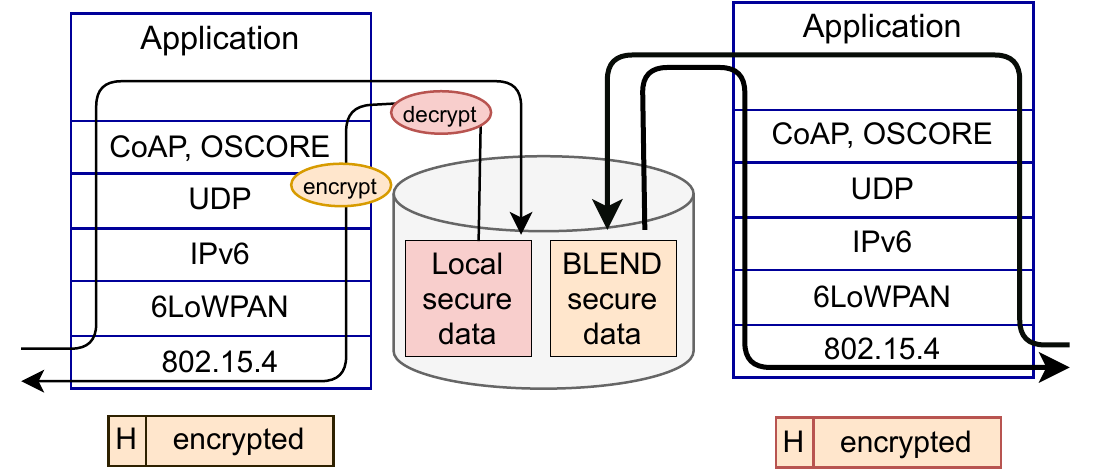}
\caption{Retrieving encrypted IoT data and securely sharing it with a remote host, with (right) an without (left) \thename}
\label{fig:AL-overview}
\end{figure}

In contrast to the active standardization work on enabling secure communication in IoT, the secure storage solutions for IoT have attracted less attention. While a custom-made local secure storage protocol can be developed, it would require new proposals on, for example, key management, choosing encryption and secure hash functions, initialization vectors, etc. Less well tested new solutions are likely to be less secure, and will require additional implementation efforts 
ultimately requiring more processing and storage resources. Most importantly, a separate secure storage solution will require additional crypto operations when an IoT data is sent to a remote host, which will increase the real-time latency. As shown in Figure~\ref{fig:AL-overview} (left), before sending a securely storage data, separate secure storage and secure communication solutions will require that the data must be first \emph{decrypted} using one set of security protocol and \emph{encrypted} again with another set of protocols. Such a solution has significant performance overhead and is infeasible for resource-constrained IoT devices. 

In this paper, we propose \thename ~that exploits novel application layer security protocols and provide combined secure storage and communication without compromising end-to-end security. \thename~does not require separate protection for storage and for communication, and the stored secure data can be shared with a remote host without any crypto operations during the transmission phase, ultimately reducing the real-time latency significantly; this is depicted in Figure~\ref{fig:AL-overview} (right). \thename~is particularly advantageous  in use cases having hard latency requirements; for example, when a drone cost-effectively collects IoT data from vast smart agriculture deployments or from remote power lines.

The main challenge in providing a combined secure storage and communication is to enable a solution that incurs minimal overhead for IoT devices, keep well-tested security properties intact, and does not compromise standard compliance and interoperability. This can be achieved by extending the use of the newly standardised OSCORE and EDHOC protocols to secure data storage. The core contributions of the paper are as follows: we \textit{(i)} extend standard based application layer security mechanisms and enable combined secure storage and secure communication; \textit{(ii)} provide an implementation of \thename ~for resource constrained devices using \theos; and \textit{(iii)} evaluate \thename ~to show its suitability for IoT.

The rest of this paper is organized as follows: related work and relevant background are presented in section \ref{sec:related} and \ref{sec:tech}, respectively;  we present a treat model in Section \ref{sec:threat}; elaborate our design in section \ref{sec:design}; provide implementation details in section \ref{sec:implement} and evaluation in \ref{sec:evaluate}; highlight security considerations in Section \ref{sec:security}; and conclude the paper in section~\ref{sec:conclusion}.

\section{Related Work}
\label{sec:related}
\subsection{Secure communication}
The area of secure communication for resource constrained devices has seen a rapid development the last decade, with the introduction of protocols targeting IoT. Early standards such as IPSec has largely been replaced with DTLS, Datagram Transport Layer Security. Still the protocol overhead is relatively large. Especially for low power radio networks where network radio packets are as small as 127 bytes, and fragmentation can cause delays and security vulnerabilities. This is limiting the usable payload for application layer sensor data down to a maximum of 51 bytes per packet, unless network specific optimizations such as 6lowPAN header compression is used, which limits the general applicability \cite{protocol-byte-overhead, RFC6282-hc, RFC7400-ghc}.
Recently new application layer protocols for secure communication have been devised which can reduce the per packet overhead, while supporting crypto algorithms suitable for constrained devices. OSCORE together with with EDHOC for key establishment have the potential to be used for PKI solutions with sufficiently low overhead for IoT. While DTLS has been shown to be feasible for PKI solutions for IoT the cost of key establishment when using standard X509 certificates is high~\cite{Indraj,PKI4IoT}.

\subsection{Secure storage}
The area of secure storage has seen much less standardisation efforts. Instead several overlapping fields are contributing to the area. Blockchain based research efforts, including \cite{BeeKeeper-blockchain, Edge-blockchain} design solutions for custom deployments, but mainly address computationally capable end devices such as cellphones or routers, and rely on custom server infrastructure. 

Another related area is research on Trusted Execution Environments, TEE, such as ARM’s TrustZone. TrustZone functionality has been used as a building block to construct secure storage for Android based devices~\cite{DroidVault}. An important area for TEE is to enable the creation of secure key storage~\cite{TrustZone-overview, SecureBlockDevice, AndroidKeystorage}. With respect to the more constrained IoT devices the TEE related efforts are complementary to our work on secure storage.

Besides the problem of secure key storage, many of the relatively lightweight cryptographic solutions used in communication protocols can be applied to any data to create a secure sensor data storage. As long as the secure storage only serves local encryption purposes, the need for standardisation is less emphasized. 

There are two previous suggestions on how to combine secure communication with secure storage, FUSION and FDTLS~\cite{FUSION, FDTLS}. The proposed designs are based on IPsec and DTLS, where promising results in terms of reduced overhead when packets are being sent are shown. An important finding is the need to optimize the storage operations with respect to the memory hardware constraints, such as to write full memory pages to reduce flash handling overhead. 

The main shortcomings of these lower layer security approaches are the following: The solutions rely on PSK, pre-shared keys. This is an outdated mode of key management, with no support for automated key management, including enrollment or revocation. Both IPsec and the DTLS version 1.2 used for the evaluations have large headers, greatly reducing the space available for sensor data when used over low power radio networks. To partly alleviate this, both designs rely on using 6lowPAN header compression, which ties the usage completely to networks where this is available. To allow new connections the protocol is side stepped in terms of removing the randomness used when generating session keys, without analyzing the security implications of this procedure, plus other minor protocol breaking tweaks. Additionally, by relying directly on IPsec or DTLS none of the conveniences offered by CoAP are available for any of the involved parties.

The conclusion is that while several works address some of the issues of secure storage of data for IoT, the existing proposals for coalesced storage and communication have serious shortcomings. We address these shortcomings with a design making use of application layer security.

\section{Necessary Background}
\label{sec:tech}
Object Security for Constrained RESTful Environments is an application-layer protocol specifically designed for IoT security~\cite{RFC8613-oscore}. It protects CoAP messages and builds upon COSE \cite{RFC8152-cose} and CBOR functionality for encryption and encoding \cite{RFC7049-cbor}. The protocol offers replay protection using sequence numbers tied to the security context. Since UDP packets might arrive out of order, the protocol uses a replay windows, such that the receiver keeps a range of currently accepted numbers. 

Ephemeral Diffie-Hellman Over COSE (EDHOC) is a proposed key exchange protocol primarily design for OSCORE~\cite{EDHOC}, and shares the usage of COSE and CBOR encoding with OSCORE. It can be used with standard X.509 certificates, or with more compact certificate formats. 
The security functionality of EDHOC is based on the SIGMA schema, from which it follows that as long as the included components keep their security guarantees, the resulting protocol will provide the desired security services~\cite{SIGMA}.

A successful EDHOC security context establishment will result in the parties agreeing on a Master Secret, a Master Salt, client and recipient IDs, and the crypto algorithms to use. With this information in place, Sender Key, Recipient Key and Common IV can be derived and saved. Once a security context is established, an endpoint is free to act both as server and client, using the same security context for both purposes~\cite{RFC8613-oscore}.

\section{Threat model and assumptions}
\label{sec:threat}
We consider scenarios where an attacker can, with some probability, get physical access to the node and probe the device permanent flash memory. We discuss both scenarios where we assume that the non-permanent memory is sufficient for key storage and scenarios where a (small) tamper protected memory area exists, which can be used for key storage. 
For communication, the Dolev-Yao threat model is applicable. An attacker can eavesdrop any communication between the involved entities, and also modify and re-send any message. As a consequence protection for replay attacks are needed, together with authentication and confidentiality services to prevent unauthorized access to any secret content.
To generate new keys and perform secure key exchange the devices must have access to a sufficiently strong random number generator.
We assume that the standards we use as building blocks are not compromised, but can offer the claimed security guarantees when used together with the recommended crypto algorithm suits.

\section{\thename: design}
\label{sec:design}

\subsection{Requirements}
The main requirement is to offer secure storage with low latency for data sending, while keeping the overall overhead low. To preserve security guarantees offered by OSCORE, as few deviations from the protocol usage as possible should be done. Preferably the receiving end of the communication should not need to take any additional steps outside of the regular protocol to receive and decrypt previously stored sensor data. In order to preserve the protocol guarantees, the initial key establishment needs to happen before packets can be precomputed and stored. 

\subsection{System building blocks}
An EDHOC implementation is needed for key establishment, but requires only standard functionality in terms of key export interfaces to create and retrieve the shared secrets used for the security context.

The OSCORE implementation needs to be augmented with handlers to enable \thename ~to precompute packets and send them unmodified at a later point in time. Practically this means allowing retrieval of the byte buffer representing the serialized OSCORE packet and ensuring there are interfaces to control the sequence numbers.

A flash storage abstraction is useful to hide hardware specific details and offer a higher level API. We propose a simple file system like API which allows reading, writing and appending data to files, which are being written out to flash. 

\begin{figure}[t]
\centering
\vstretch{.9}{\includegraphics[width=0.6\columnwidth]{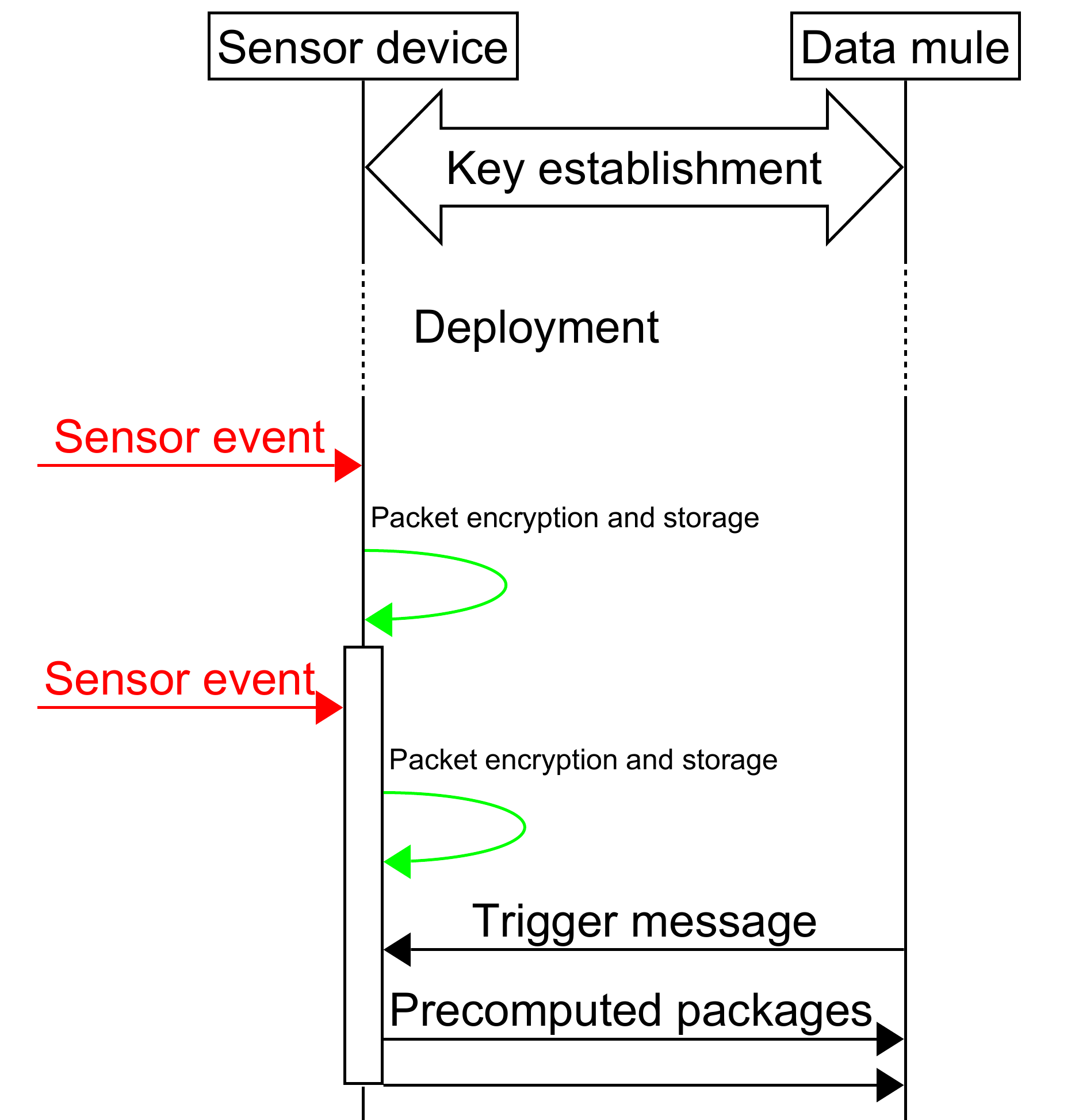}}
\caption{\thename ~overview. An initial key establishment is done before deployment, can be redone later given EDHOC support. Sensor data is encapsulated into precomputed packets, and securely stored until a connection with a data mule, or any other secure endpoint, is available}
\label{fig:AL-fusion}
\end{figure}

\begin{figure*}[t]
\centering
\includegraphics[width=0.71\textwidth]{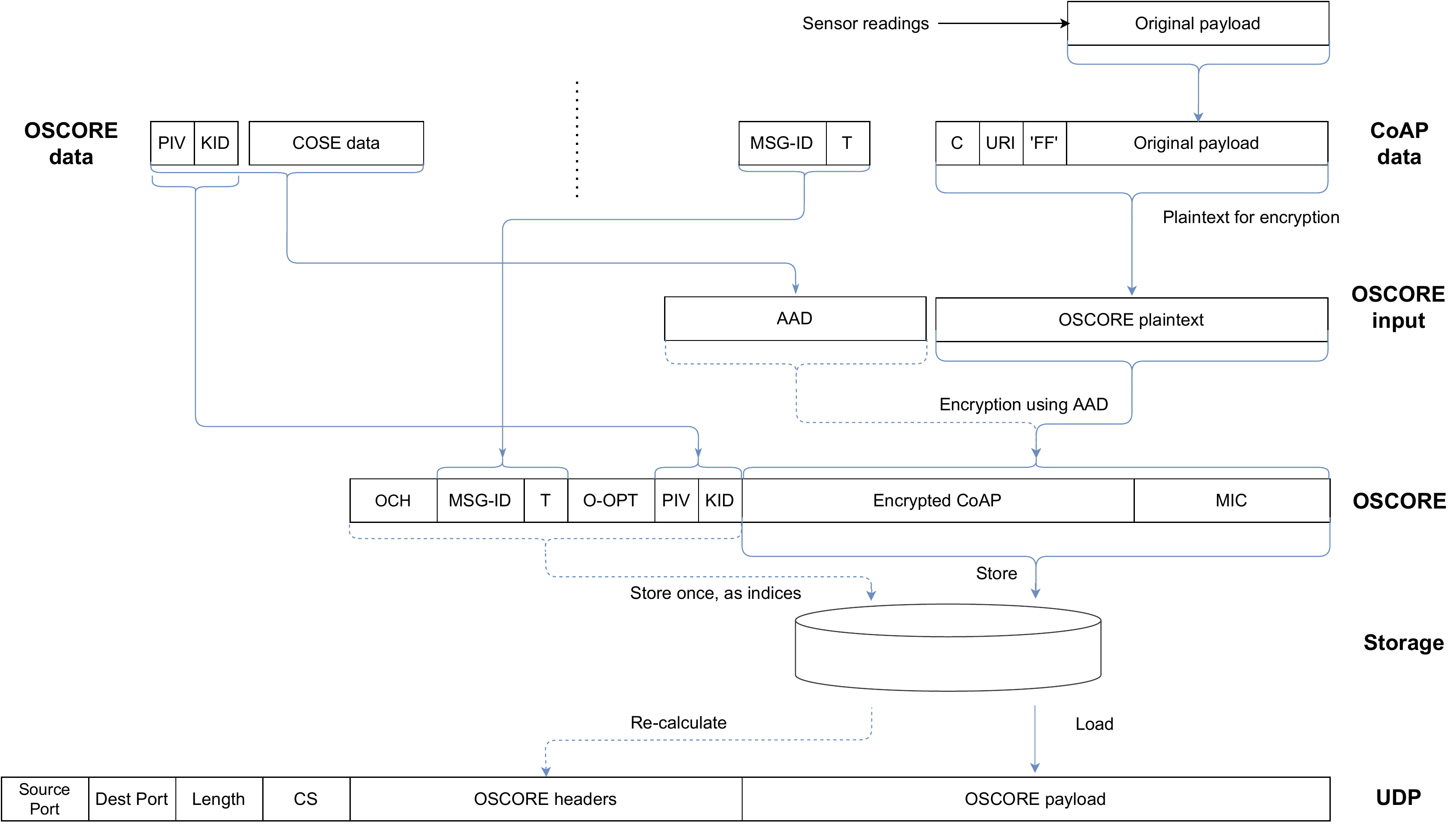}
\caption{The relation between the layers when using \thename}
\label{fig:AL-fusion-packets}
\end{figure*}

\subsection{SecureStorage lifecycle and message flow}
The figure \ref{fig:AL-fusion} illustrates the main events relevant to secure storage operations. After the key establishment both parties have established a secure context, which allows them to act as both clients and servers.

The sensor can thereafter be deployed, and start sensing. Depending on the data generation rate and storage policy, a number of sensor readings might be compiled as the payload for one CoAP packet. The packet is encrypted as a ready to send OSCORE packet and stored onto flash. 

When the communication link is ready, for instance in the form of a data mule, a trigger command is sent.
The trigger message is a CoAP request, protected with the same OSCORE security context as has been previously established through EDHOC. Hence the correct decoding of the trigger message serves both to authenticate the data mule, and to authorize it for accessing the sensor data. 
The command will cause the sensor device to start sending the stored packages. In sending the stored packages, the sensor device acts as a client. This allows the device to control the sequence numbers included in the packets, reflecting the sequence numbers of the stored packets.
To prevent an attacker from stopping a data transfer simply by blocking the trigger message, we require the device to reply with a short no data message in case there is no sensor data to send.

\subsection{Storage overhead trade-offs}
The amount of data that the sensor device needs to store locally depends on both the sensor data generation rate and the frequency of data collection from the outside. For sensor devices deployed in low power radio networks, the least amount of overhead is achieved if payloads corresponding to full 802.15.4 frame sized packets are precomputed and stored. If the sensor data generation rate is sufficiently small, the node would need to temporarily store unencrypted sensor data until the amount corresponding to a full payload is gathered. If no temporary plaintext storage is considered acceptable, the device must create encrypted packets for each individual sensor reading.

\begin{table}[ht]
\caption{Plaintext data needed to prepare the CoAP packets used in \thename}
\small
\centering
\begin{tabular}{|l l l l|} 
\hline
\textit{Type}	&	\textit{Size}, byte	&	\textit{Example}	&		\\
\hline\hline							
Version \& type	&	1	&	'40'	&	ver.1, confirmable	\\
\textbf{Code}	&	1	&	'02'	&	POST	\\
\textit{Message ID}	&	2	&	'4A 84'	&	$<$any id$>$	\\
\textit{Token}	&	1	&	'84'	&		\\
\textbf{URI path \& len}	&	1+path len	&	'b0'	&		\\
\textbf{Payload marker}	&	1	&	'FF'		&	\\
\textbf{Payload}	&	6--56	&	\multicolumn{2}{l|}{$<$binary data$>$}\\
\hline\hline							
\textbf{To be encrypted}	&	\multicolumn{2}{l}{\textit{$\ge$ 3+payload len}}			&		\\
\hline
\end{tabular}
\label{table:coap_layer}
\end{table}

\begin{table}[ht]
\caption{Data contained in the OSCORE packets}
\small
\centering
\begin{tabular}{|l l l l|} 
\hline
\textit{Type}	&	\textit{Size}, byte	&	\textit{Example}	&		\\
\hline\hline							
Version \& type	&	1	&	'40'	&	ver.1, confirmable	\\
Code	&	1	&	'02'	&	POST	\\
Message ID	&	2	&	'4A 84'	&	from CoAP	\\
Token	&	1	&	'84'	&		from CoAP\\
OSCORE flag	&	2	&	'93 09'	&		\\
Partial IV	&	1--	&	'13'	&	$=$ sequence no	\\
Key ID	&	1	&	'42'	&	$=$ sender id	\\
Payload marker	&	1	&	'FF'	&		\\
Encrypted payload	&	9--59	&	\multicolumn{2}{l|}{$<$encrypted CoAP$>$}			\\
MIC	&	8	&		&		\\
\hline\hline							
\textit{Packet length}	&	\multicolumn{3}{l|}{\textit{$\ge$ 21+original sensor data payload len}}					\\
\hline
\end{tabular}
\label{table:oscore_layer}
\end{table}

\begin{table}[ht]							
\caption{Previous state-of-art, data contained in a DTLS record packet}	
\small							
\centering							
\begin{tabular}{|l l l|}							
\hline							
\textit{Type}	&	\textit{Size}, byte	&	\textit{Example}			\\
\hline\hline							
Content type	&	1	&	'17'~~~~~$=$ application data			\\
Version	&	2	&	'FEFD'~$=$ DTLS 1.2			\\
Epoch	&	2	&	'0001'			\\
Sequence number	&	6	&	'0000 0000 0001'			\\
Length	&	2	&				\\
Initialization vector	&	8	&				\\
Encrypted payload	&	6--51	&	$<$encrypted raw sensor data$>$			\\
MIC	&	8	&				\\
\hline\hline							
\textit{Packet length}	&	\multicolumn{2}{l|}{\textit{$\ge$ 29+original sensor data payload len}}					\\
\hline							
\end{tabular}							
\label{table:dtls_layer}							
\end{table}

\subsection{Relation between the layers}
An detailed illustration of the relation between the layers is shown in figure~\ref{fig:AL-fusion-packets}. In the following we explain the data needed to be included and processed.

\subsection{CoAP packet creation}
While the CoAP protocol is versatile, with a range of packet options, we are here interested in a meaningful minimal subset needed to precompute sensor data packets. The table~\ref{table:coap_layer} shows the minimal plaintext data needed to create a CoAP packet, ready to be encrypted for secure storage. In italics are the fields that will be moved to the OSCORE packet. In bold are the fields will be protected through encryption. An observation is that the length of the destination URI directly adds to the packet overhead, but unless otherwise required the empty root path can be used as a valid destination URI.

\subsection{OSCORE packet creation}
Given an existing security context and the CoAP packet information, \thename ~can encrypt the CoAP payload together with the sensitive header fields, and calculate the correct OSCORE headers. The missing dynamic information needed is the sender sequence number. The sequence number is used as the basis for the partial initialization vector, or Partial IV in COSE terms. The sender ID is used as key ID. ('PIV' and 'KID' in figure \ref{fig:AL-fusion-packets}.) These two items, together with static COSE information on the algorithm used, are used to form the additional authenticated data, AAD, used in encryption. The two items are also used for calculating a nonce used in encryption, and finally they are included in plaintext in the OSCORE packet header. 

The table~\ref{table:oscore_layer} shows the data present in the resulting OSCORE header. Starting from the original sensor data, the minimal total overhead is 21 bytes. For sequence numbers in the range 255--65535 an extra byte is needed, etc. This flexible sizing is in contrast to the older DTLS standard (shown in table \ref{table:dtls_layer}), where a fixed field of 6 bytes is allocated regardless of the currently needed size.

\subsection{Storage of precomputed packets}
For systems with fast flash memory operations, or where energy is of less concern, the prepared OSCORE packet can be saved directly. Where flash operations are slow or energy efficiency is paramount, the OSCORE packet header information can be stored once for a whole series of sensor data packets. Since all the dynamic fields; the message ID, token and sequence number can be assigned in a predictable increasing manner, storing and later retrieving the starting points for the first packet header is sufficient to recalculate the following packet headers. It is this optimized procedure which is shown in figure~\ref{fig:AL-fusion-packets}.

\subsection{UPD alternative}
Also UPD headers could be precomputed, and the entire UDP databuffer could be stored for minimal processing at the time of sending. Precomputing UDP packets requires the source and destination ports to be known beforehand. A more important drawback is the increased storage overhead, since the UDP headers add another 8 bytes to each precomputed packet, which needs extra time for storage and retrieval.

\subsection{Key management}
\thename~relies on devices being able to establish new security contexts upon need. To create a new context with the same endpoint, OSCORE allows existing master secret data to be reused, making the context derivation computationally cheap. This can be used to keep the context sequence number bounded by a fixed length. For key establishment we propose EDHOC to be used. EDHOC offers relatively low overhead while supporting PKI solutions. Low overhead is achieved through using certificate reference based key establishment. This requires relevant certificates to have been securely distributed at an earlier point in time. Certificate distribution is out of scope for this work, but in contrast to solutions based on shared secrets, certificates are meant to be openly shared and could be distributed from any trusted endpoint.
\subsubsection{Planned secure context updates}
If the data collection endpoint is replaced, the sensor device needs to establish a new security context with the new endpoint. For a planned update, a notice can be communicated ahead of time. Depending on the deployment scenario, this might simply be a message to initiate a full new key establishment, which allows the sensor to immediately start using the new context for sensor data storage. For extreme deployments with very limited connectivity and data mules, it might be a notice send during the last data collection round where the old data mule is active. Unless the key establishment can be relayed at that time, the sensor device has to temporarily resort to local storage encryption until a new security context is in place.
\subsubsection{Unplanned secure context losses}
For cases when the data connection endpoint loses the security context, or is lost all together, the following round of data collection with a not previously used endpoint requires re-keying. Any data that has been stored locally using the old security context need to be decrypted by the IoT device and encrypted again.

For IoT devices with access to tamper resistant nonvolatile memory that can be used for key storage, they can store the shared secret data established through the key exchange such that they can recover the security context in case of temporal power losses or restarts.

An IoT device without a secure permanent key storage wants to minimize the storage of security sensitive data to a minimum. Hence there is a risk of losing vital parts of the security context, in case of power losses and unplanned restarts. In the case of context losses the previous stored precomputed sensor data packets become opaque to the device. The stored data can still be sent to the endpoint which has access to the security context and is able to decipher the encrypted packets. Depending on the deployment, the device might report its situation and request a new authentication through a new key exchange before sending the packets from the old security context. In this case the receiving endpoint must keep both contexts in parallel. Alternatively, the setup can be done to allow the IoT device to interpret an incoming message it cannot decipher as the expected trigger message, if it is recovering from a security context loss.

\subsubsection{Proximity to endpoint}
Since EDHOC offers true end-to-end protection it can be used to establish a security context with any reachable remote endpoint, even behind proxies. 

\subsection{Re-sending and multiple receivers}
The usage of precomputed sensor data packets does not affect resending that happens on lower layers. Lower layer resending will depend on the deployment scenario and radio configuration. As long as a packet has not been received by the other end, the receive window used for the replay detection by the recipient remains unchanged. If on the other hand data has been received the same packet can no longer be resent, 
as the encryption is affected by the sequence number.
For scenarios where either the same receiver wants the same data item more than once, or where multiple receivers are interested in the same data item, extensions of the keying schema must be done. To handle multiple receivers there are proposals for OSCORE group communication, which could be part of an extended secure storage solution \cite{I-D.ietf-core-oscore-groupcomm}.

\section{Implementation}
\label{sec:implement}
We have implemented \thename ~in C as a module for the Contiki NG embedded OS~\cite{embedded-os}, that can be adapted for other available operating systems such as Zephyr~\cite{ZephyrSite}. The \thename ~implementation contains the needed OSCORE libraries, including COSE and CBOR encoding and decoding.

For the basic crypto operations we have reused functionality from the crypto libraries available in Contiki NG, which offers partial crypto operation hardware acceleration for selected target platforms.

\paragraph{Secure communication}
The secure communication part of \thename ~is build using the OSCORE libraries available in an experimental version of Contiki NG, plus our EDHOC implementation for key establishment. To allow reusability of the available code for confirmable CoAP transactions we include a CoAP token in the packets.

\paragraph{Secure storage}
The storage part is built on top of the Coffee file system for Contiki, offering a file abstraction for interacting with underlying flash memory. When the optimized packet storage method is used, the dynamic header information needed to recalculate the full headers is recorded at the start of new files, followed by the encrypted part of the packets. The specifics of flash memory block sizes and optimal amount of data to write per file depends on target hardware. 

\paragraph{Key management and crypto algorithms}
We have implemented EDHOC which is used to establish shared secrets, and based on them derive security contexts. Both the EDHOC and the OSCORE standards are flexible in terms of supporting multiple crypto suits. Our implementation is focused on the mandatory SHA-256 for HKDF, HMAC-based Extract-and-Expand Key Derivation Function, and the most commonly used symmetric crypto, AES-CCM-16-64-128. While AES-CCM is a block cipher, it does not require padding of the resulting ciphertext. As a result the length of ciphertext is always the length of the plaintext, plus 8 byte MIC, message integrity code.

\section{Evaluation}
\label{sec:evaluate}
We use a quantitative experimental research methodology where we evaluate the impact of one particular variable while keeping other parts of the system setup static, to correctly attribute the performance variations. In the following we present the relevant micro benchmarks illustrating the system performance and overhead.

\subsection{Experimental setup}
For the hardware experiments we use Zolertia Firefly nodes, a platform using TI CC2538 ARM Cortex M3 micro-controllers~\cite{Zolertia:Firefly}. The nodes are equipped with 32 KB RAM, 512 KB flash, a 2.4GHz 802.15.4 radio for communication and support for hardware acceleration of crypto operations.

\begin{figure}[t]
\centering
\includegraphics[width=0.75\columnwidth]{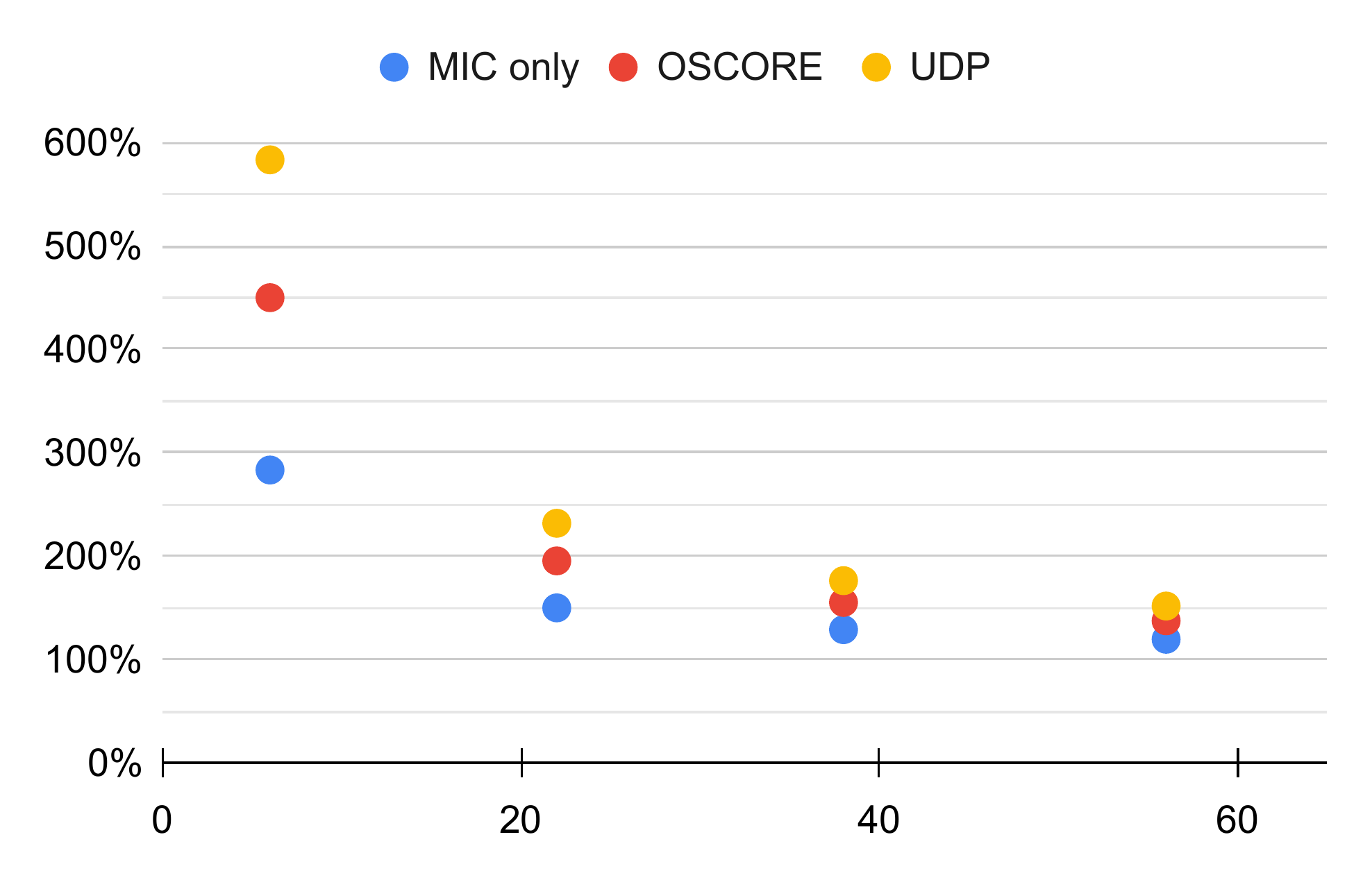}
\caption{Storage size relative to the original sensor data size in bytes, for different sensor payloads and storage options}
\label{fig:SS-storage-oh}
\end{figure}

\subsection{Storage overhead}
The packet storage overhead for different sensor data payloads is shown in figure \ref{fig:SS-storage-oh}. Storing a ready to send OSCORE packet induces an overhead of 21 bytes, using the configuration presented in tables \ref{table:coap_layer} and \ref{table:oscore_layer}. If instead a complete UDP packet is stored, the per packet overhead is 29 bytes. If only the starting points for dynamic header data counters are stored once, the overhead quickly shrinks to close to 3 extra CoAP bytes, plus the 8 byte MIC. For our system tests we use file append functionality for storing packets, such that the storage cost of the 6 bytes needed for header recalculations are amortized over 25 precomputed packets.

Depending on the initial sensor data size the resulting storage overhead ranges from 20\% for the 56 byte sensor data packets using optimized storage, all the way up to close to 600\% for 6 byte sensor data while storing full UDP packets. 

In the following experiments the optimised version where needed header data is stored once is used.

\begin{figure*}[!ht]
\centering
 \begin{subfigure}{0.45\textwidth}
 \includegraphics[width=0.78\columnwidth]{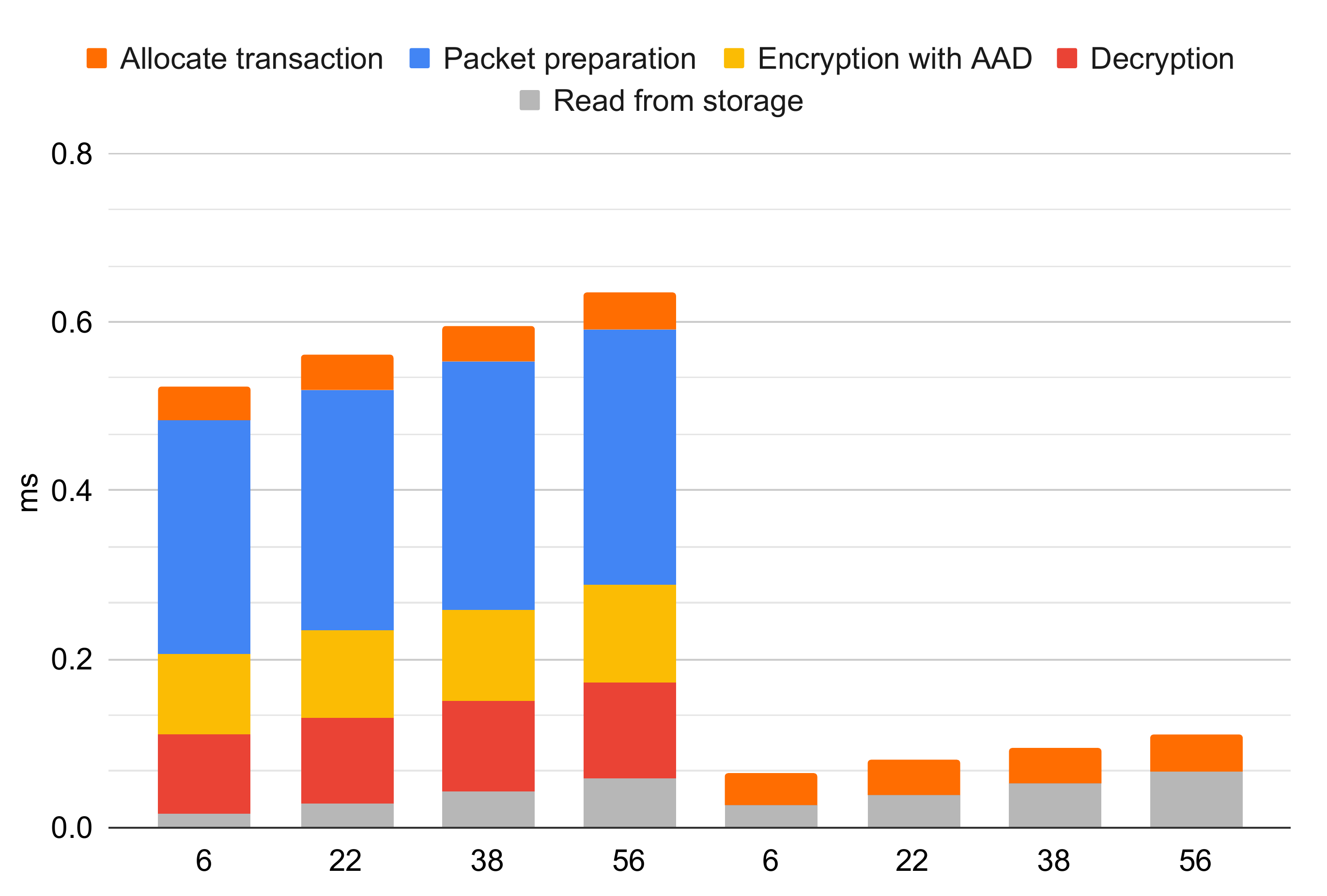}
 \caption{with hardware acceleration}
 \label{fig:SS-prepare-hw}
 \end{subfigure}
\begin{subfigure}{0.45\textwidth}
\includegraphics[width=0.78\columnwidth]{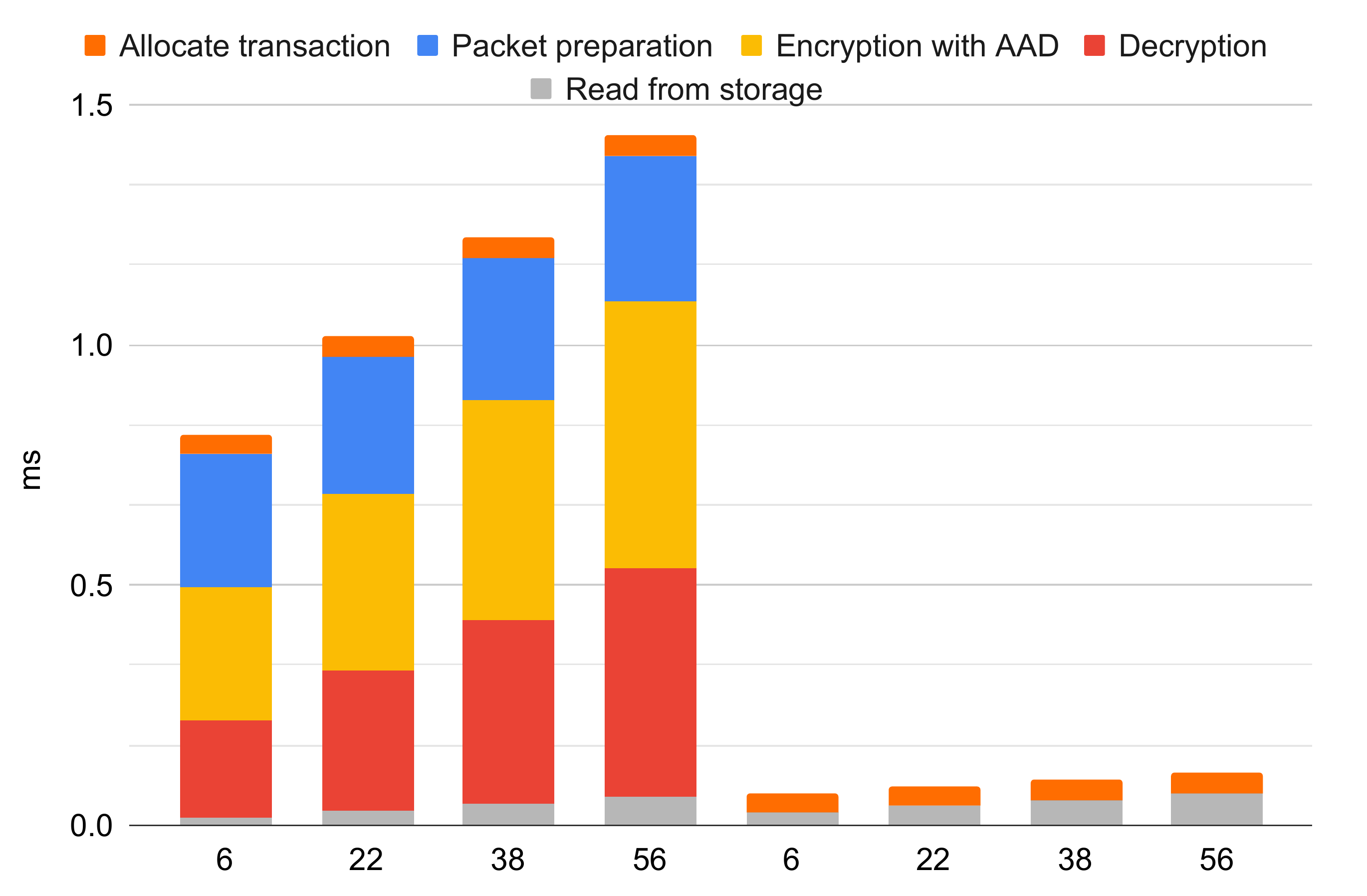}
\caption{No hardware acceleration}
\label{fig:SS-prepare-sw}
\end{subfigure}
\caption{Latency for packet preparation, with local security and while using \thename}
\end{figure*}

\subsection{Latency to get data ready for sending}
When the device gets a request to report recorded sensor readings, if local security is used, it needs to read the data from flash, decrypt it with the local key and prepare it for sending. If \thename ~is used, the operations needed are reading from flash and, optionally, packet transaction allocation. The two cases are illustrated in figure~\ref{fig:AL-overview}. The total time needed is shown in figure~\ref{fig:SS-prepare-hw}, for when hardware acceleration is available, and in figure~\ref{fig:SS-prepare-sw} without hardware acceleration.

With hardware acceleration \thename ~performs around 0.5 $ms$ faster compared with the local security solution. The resulting remaining latency when retrieving a stored packet, recalculating header information and allocating a CoAP transaction is is only between 65 $\mu$$s$ and 110 $\mu$$s$.

Without hardware acceleration, with all cryptographic operations done in software, the latency savings are between 0.75 $ms$ and 1.36 $ms$ per packet compared with the local security solution. 

\begin{figure*}[!ht]
\centering
\begin{subfigure}{0.45\textwidth}
\includegraphics[width=0.78\columnwidth]{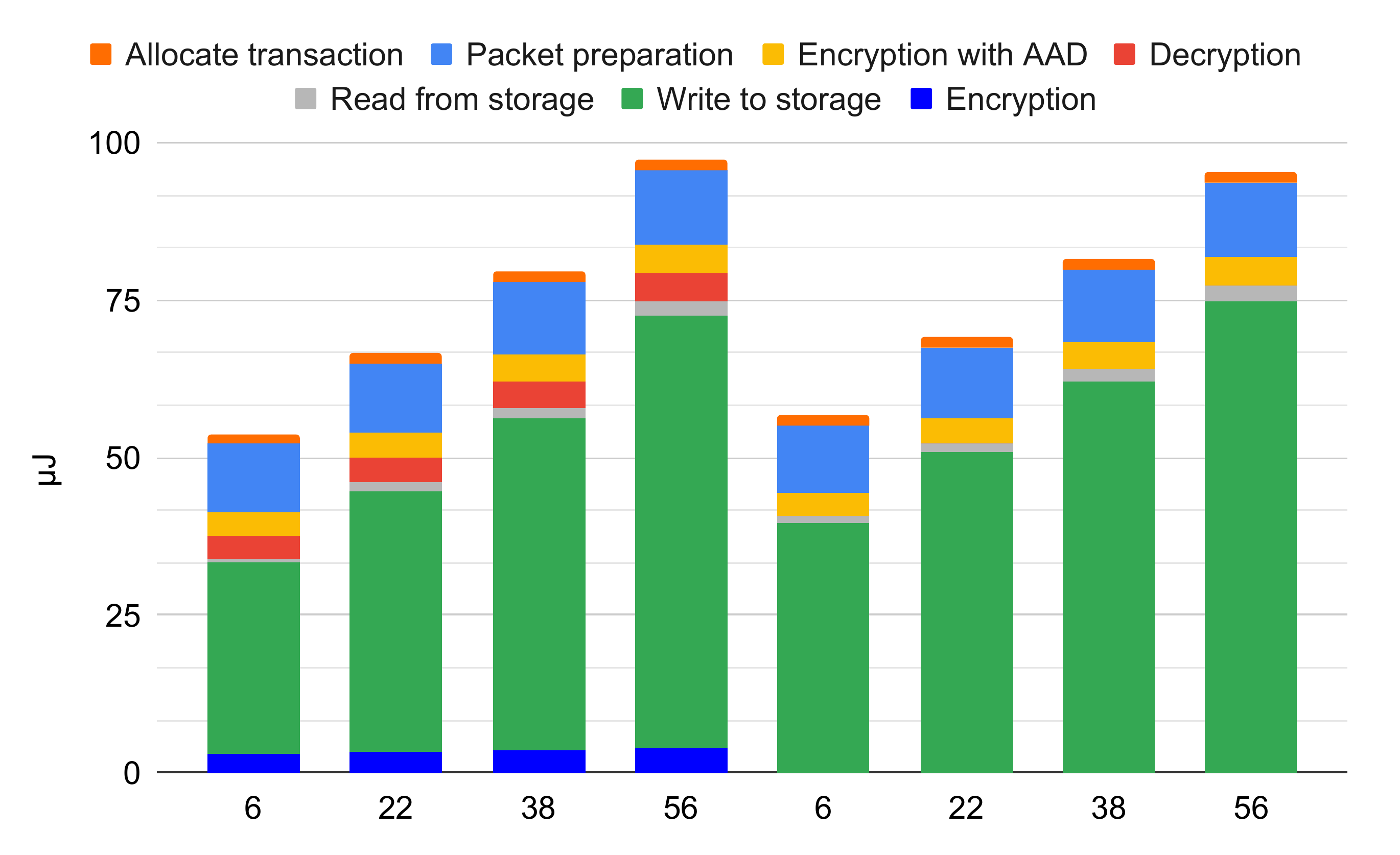}
\caption{With hardware acceleration}
\label{fig:SS-energy-hw}
\end{subfigure}
\begin{subfigure}{0.45\textwidth}
\includegraphics[width=0.78\columnwidth]{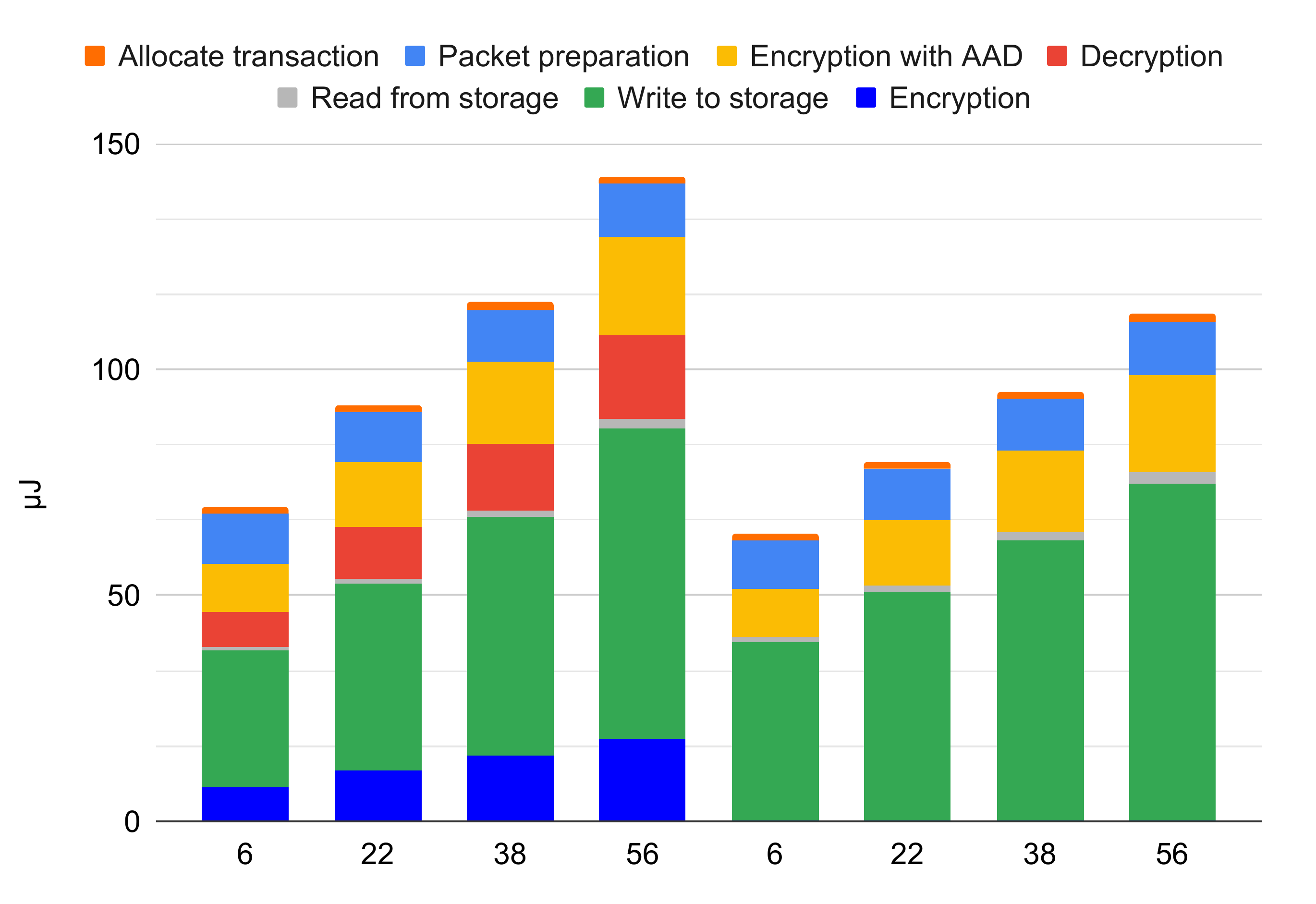}
\caption{No hardware acceleration}
\label{fig:SS-energy-sw}
\end{subfigure}
\caption{Total energy needed for sensor data operations, with local security while using \thename}
\end{figure*}

\subsection{Total energy usage}
Using the fine grained timer system in Contiki NG we measure the time spent for relevant system operations. This makes it possible to calculate the consumption based on current and voltage levels from the CC2538 hardware datasheets \cite{Zolertia:Firefly}. We use the specified maximum peak current for writing, which means we present the absolute upper bound of energy usage for the flash operations. 

The total energy usage is shown in figure~\ref{fig:SS-energy-hw} and \ref{fig:SS-energy-sw}, with and without the usage of crypto hardware acceleration. Using the crypto hardware acceleration the differences are small. Due to relatively slow storage write operations, also a small increase in storage needs can offset crypto savings. Without crypto hardware acceleration, \thename~saves energy for all sensor data sizes.

The conclusion is that \thename ~performs at least on par with the local storage solution in terms of total energy usage, with a clear advantage for all cases where the crypto operations constitutes a larger proportion of the total work done.

\subsection{Key establishment and comparison with DTLS}
\subsubsection{Key establishment}
Using the reference based key establishment option in EDHOC we are able to perform a key establishment using only 284 bytes of application layer data. With DTLS 1.2 and the ECDHE-ECDSA cipher suit corresponding operations use 1.65 kB.
This is largely due to lengthy ASN.1 encodings and the need to send full certificates in the handshake. The numbers are based on IoT profiled certificates of only 315 bytes for both parties in the exchange.
\subsubsection{Packet encryption and overhead}
The AES crypto used for OSCORE corresponds to what is commonly used also for DTLS 1.2 in IoT devices. This means the overhead from the crypto operations are directly comparable. An obvious benefit of switching to an OSCORE based solution is the reduced packet overhead. Using the DTLS AES128-CCM8 cipher produces the packet overhead figures given in table \ref{table:dtls_layer}, to be compared with numbers for OSCORE in table \ref{table:oscore_layer}. An OSCORE solution using CoAP saves eight bytes even compared with a DTLS session without CoAP, used to transport raw UDP data. If instead also DTLS is used to provide a CoAPs session, encryption will be performed on the whole CoAP layer packet, which reduces the maximum usable sensor data payload with six more bytes, down to 45 bytes.
\begin{figure}[t]
\centering
\includegraphics[width=0.75\columnwidth]{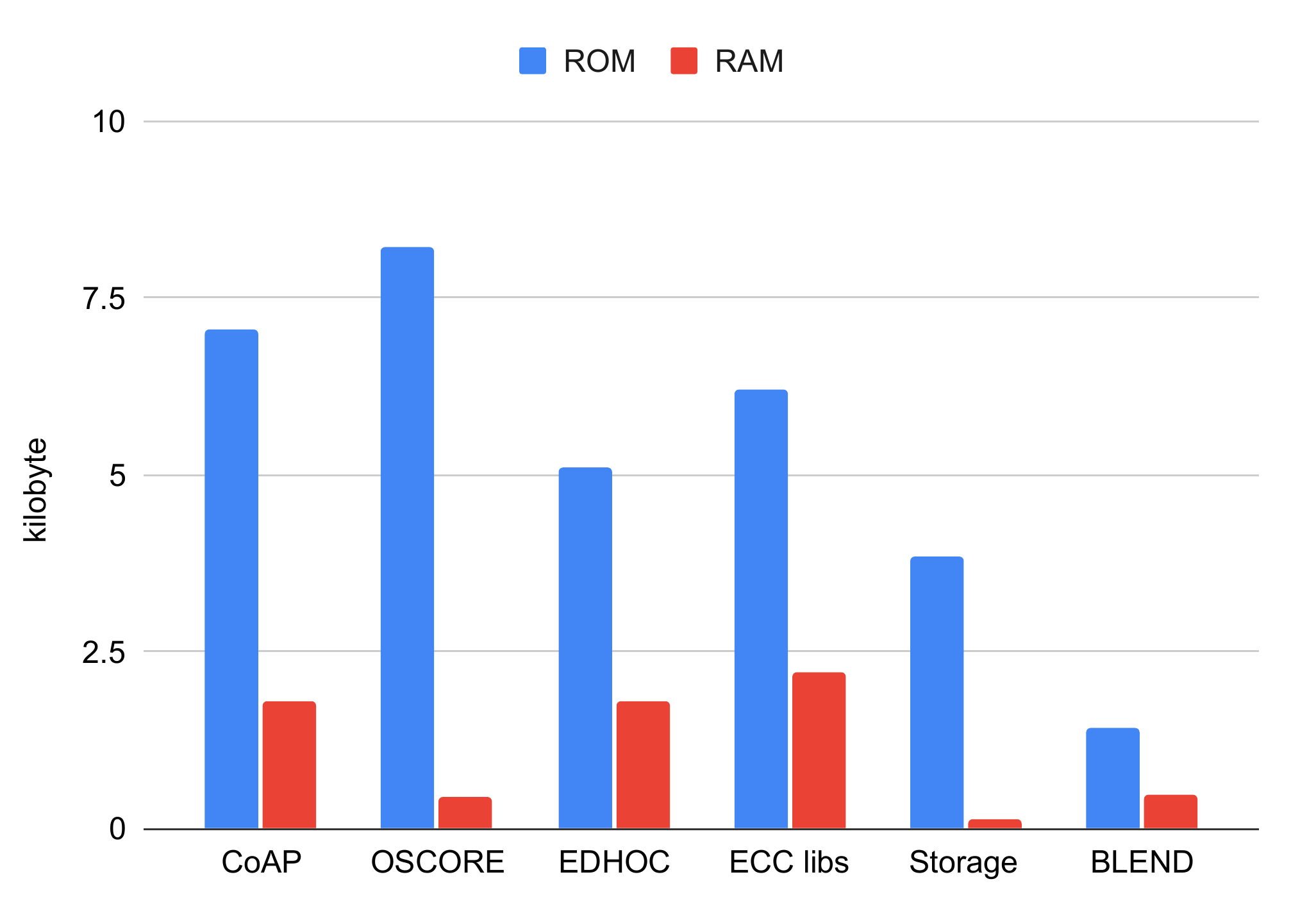}
\caption{Memory usage by \thename ~and related components}
\label{fig:SS-memory}
\end{figure}

\subsection{Memory requirements}
The \thename ~implementation requires 1.5 kB of ROM and a little less than 0.5 kB of RAM. Figure \ref{fig:SS-memory} shows the comparison with the related components in the configuration used. Compared with the size of the total Contiki NG firmware used for the evaluation of around 60 kB ROM and 13 kB of pre-allocated RAM, the \thename ~only contributes to 2.5\% of the ROM and 3.8\% of the RAM.

The numbers shown are when there is memory allocated for two parallel security contexts. Each additional security context adds 143 bytes of RAM.

\section{Security Considerations}
\label{sec:security}
When a protocol designed for securing communication is reused to also protect data at rest, it is important to validate that protocol assumptions are still applicable. This includes the amount of data that can be protected. The OSCORE protocol is designed to allow theoretical maximum sequence numbers up to $2^{40}$-$1$, but for actual implementations the number will be lower. The implementation used in our evaluation allows sequence numbers up to 4.3 billion. Using 48 byte sensor data packets, to ensure no fragmentation, this corresponds to covering more than 200 GB of data without resetting the sequence counter. This is not a limiting factor for the resource constrained IoT scenarios considered.

The EDHOC protocol relies heavily on the availability of a secure random number generator. For devices with less strong random generators there are proposals on how to incorporate more random material to improve generator quality~\cite{RFC8937-improve-randomness}.

If the same master secret data is used to generate multiple secure sessions, forward secrecy is no longer guaranteed \cite{RFC8613-oscore}. If the long-term secret is leaked, data from previous sessions risk being exposed. This means the multiple session feature should only be used when the risk that previous communication has been eavesdropped is either neglectable, or if the old data no longer is considered secret.

Our only proposed deviation from existing protocol compliance is the suggestion that an IoT device that has lost its secure session could be allowed to send its stored encrypted data without performing mutual authentication and establishing a new secure session. This could enable an attacker to trick the device into sending data, but which the attacker will not be able to decipher, as long as the protocol is not otherwise compromised. To prevent the data from getting lost, the device should keep the data until it has been properly acknowledged through a new security context.

The concerns regarding secure key storage are applicable for any local secure storage solution as well. The local storage needs either a long time key stored in persistent memory, or it needs a secure key management protocol of its own. 

\section{Conclusion}
\label{sec:conclusion}
We have shown that the new application layer security standard, OSCORE, can be integrated with an IoT storage system, which makes it possible to provide a secure data storage service without compromising any communication security properties or the standard compliance. Our solution, \thename ~drastically reduces the latency for sending stored IoT data compared with a local secure storage solution. When combined with EDHOC for performing secure key exchange and establishing the needed security context, \thename ~ enables a resource efficient way to achieve a complete secure storage and communication solution for IoT.

\section*{Acknowledgment}
This research is partially funded by the Swedish SSF Institute PhD grant and partly by the EU H2020 ARCADIAN-IoT (Grant ID. 101020259), the ITEA3 Smart, Attack-resistant IoT Networks (Project ID: P123800021) and the H2020 CONCORDIA (Grant ID: 830927) projects.

\bibliographystyle{IEEEtran}
\bibliography{IEEEabrv,ref}

\end{document}